\documentclass[conference]{IEEEtran}
\usepackage{cite}
\usepackage{amsmath,amssymb,amsfonts}
\usepackage{algorithmic}
\usepackage{graphicx}
\usepackage{textcomp}
\usepackage{xcolor}
\usepackage{booktabs}
\usepackage{tabularx}
\usepackage{multirow}
\usepackage{multicol}
\usepackage{placeins}
\usepackage[normalem]{ulem}
\usepackage{array}
\def\BibTeX{{\rm B\kern-.05em{\sc i\kern-.025em b}\kern-.08em
    T\kern-.1667em\lower.7ex\hbox{E}\kern-.125emX}}

\PassOptionsToPackage{hyphens}{url}\usepackage{hyperref}
\usepackage{scalerel}
\usepackage{tikz}
\usepackage{multirow}
\usepackage{xspace}
\usepackage{siunitx}
\usetikzlibrary{svg.path}

\definecolor{orcidlogocol}{HTML}{A6CE39}
\tikzset{
  orcidlogo/.pic={
    \fill[orcidlogocol] svg{M256,128c0,70.7-57.3,128-128,128C57.3,256,0,198.7,0,128C0,57.3,57.3,0,128,0C198.7,0,256,57.3,256,128z};
    \fill[white] svg{M86.3,186.2H70.9V79.1h15.4v48.4V186.2z}
                 svg{M108.9,79.1h41.6c39.6,0,57,28.3,57,53.6c0,27.5-21.5,53.6-56.8,53.6h-41.8V79.1z M124.3,172.4h24.5c34.9,0,42.9-26.5,42.9-39.7c0-21.5-13.7-39.7-43.7-39.7h-23.7V172.4z}
                 svg{M88.7,56.8c0,5.5-4.5,10.1-10.1,10.1c-5.6,0-10.1-4.6-10.1-10.1c0-5.6,4.5-10.1,10.1-10.1C84.2,46.7,88.7,51.3,88.7,56.8z};
  }
}

\newcommand\orcidicon[1]{\href{https://orcid.org/#1}{\mbox{\scalerel*{
\begin{tikzpicture}[yscale=-1,transform shape]
\pic{orcidlogo};
\end{tikzpicture}
}{|}}}}

\newcommand{\eg}{e.g.,\xspace}
\newcommand{\idest}{i.e.,\xspace}

\newcommand{\comment}[1]{\ignorespaces}

\begin{document}

\title{Benchmarking Adaptative Variational Quantum Algorithms on QUBO Instances}


\author{\IEEEauthorblockN{1\textsuperscript{st} Gloria Turati}
\IEEEauthorblockA{
\textit{Politecnico di Milano}\\
Milano, Italy \\
gloria.turati@polimi.it}
\and
\IEEEauthorblockN{2\textsuperscript{nd} Maurizio {Ferrari Dacrema}} 
\IEEEauthorblockA{\textit{Politecnico di Milano}\\
Milano, Italy \\
maurizio.ferrari@polimi.it}
\and
\IEEEauthorblockN{3\textsuperscript{rd} Paolo Cremonesi}
\IEEEauthorblockA{
\textit{Politecnico di Milano}\\
Milano, Italy \\
paolo.cremonesi@polimi.it}
}

\maketitle

\begin{abstract}

In recent years, Variational Quantum Algorithms (VQAs) have emerged as a promising approach for solving optimization problems on quantum computers in the NISQ era. However, one limitation of VQAs is their reliance on fixed-structure circuits, which may not be taylored for specific problems or hardware configurations. A leading strategy to address this issue are Adaptative VQAs, which dynamically modify the circuit structure by adding and removing gates, and optimize their parameters during the training. Several Adaptative VQAs, based on heuristics such as circuit shallowness, entanglement capability and hardware compatibility, have already been proposed in the literature, but there is still lack of a systematic comparison between the different methods. In this paper, we aim to fill this gap by analyzing three Adaptative VQAs: Evolutionary Variational Quantum Eigensolver (EVQE), Variable Ansatz (VAns), already proposed in the literature, and Random Adapt-VQE (RA-VQE), a random approach we introduce as a baseline. In order to compare these algorithms to traditional VQAs, we also include the Quantum Approximate Optimization Algorithm (QAOA) in our analysis. We apply these algorithms to QUBO problems and study their performance by examining the quality of the solutions found and the computational times required. Additionally, we investigate how the choice of the hyperparameters can impact the overall performance of the algorithms, highlighting the importance of selecting an appropriate methodology for hyperparameter tuning. Our analysis sets benchmarks for Adaptative VQAs designed for near-term quantum devices and provides valuable insights to guide future research in this area.
\end{abstract}


\section{Introduction}

Adaptative Variational Quantum Algorithms (VQAs) have recently emerged as a leading strategy to overcome the limitations of traditional VQAs\cite{VQAs}. Adaptative VQAs are quantum algorithms which dynamically modify the structure of the circuit, with the goal of creating ansatzes that are tailored to the specific problem being solved and the available hardware.
The common approach adopted consists in defining a pool of gates and some rules which describe how these gates should be chosen and added to the circuit at each iteration.
The rules are inspired by the desirable properties of a quantum circuit, such as circuit shallowness, low number of noisy gates, and appropriate number of cnot gates for entanglement generation.
In Adaptative VQAs it is possible to distinguish two optimization processes. The first aims to create the structure of the circuit by adding and removing gates, while the second has the goal of optimizing the parameters of the gates (\eg angles of rotation gates).
Both structure and parameter optimization aim to produce the best circuit by minimizing their respective loss functions, which encode the information on the problem we want to solve and the properties desired for the circuit.

Several Adaptative VQAs have already been proposed in the literature. 
A first category includes ADAPT-VQE\cite{fermionic-adapt-VQE}, qubit-ADAPT-VQE\cite{qubit-adapt-VQE}, and qubit-excitation-based-ADAPT-VQE\cite{QEB-adapt-VQE}, adaptative versions of the Variational Quantum Eigensolver (VQE)\cite{VQE1, VQE2}, which are designed for the same purpose (\idest finding the ground state of a Hamiltonian operator), but do not rely on a fixed-structure circuit. Specifically, they start with an initial ansatz and dynamically add gates from a pool that depends on the specific chemistry problem being solved.
There also exists an adaptative version of the Quantum Approximate Optimization Algorithm (ADAPT-QAOA)\cite{adapt-QAOA}, which, at each iteration, chooses the most appropriate mixer for the following layer of the circuit.
Other Adaptative VQAs are based on a genetic approach\cite{MoG-VQE, genetic, EVQE}, while others develop variational ansatzes using Machine Learning techniques\cite{cincio, QAS}.
Finally, some algorithms employ strategies like gate simplification in order to reduce the circuit depth\cite{VAns}.

As there are multiple well-motivated methods, it is challenging to determine the best approach to solve a specific problem.
Moreover, several studies have already benchmarked the performance of traditional VQAs\cite{bench_vqa1, bench_vqa2} and the variants of ADAPT-VQE\cite{bench_adapt_vqe1, bench_adapt_vqe2}. However, a comprehensive comparison of different types of adaptative VQAs applied to the same problem with the goal of assessing their effectiveness compared to fixed-structure methods is currently lacking in the literature.
Therefore, in this work we aim to establish some benchmarks applying three Adaptative VQAs and QAOA to different instances of Quadratic Unconstrained Binary Optimization (QUBO) problems. 
Specifically, in our analysis we want to assess: (i) the efficacy of adaptative algorithms to identify a good structure for the circuit, (ii) the effectiveness of Adaptative VQAs compared to methods that use a fixed structure, (iii) the impact of the hyperparameters on the performance of the algorithms.

\section{Algorithms Benchmarked}

The goal of this study is to evaluate the efficacy of adaptative algorithms in optimizing quantum circuits. For this reason, we choose the following algorithms: Evolutionary Variational Quantum Eigensolver (EVQE)\cite{EVQE} and Variable Ansatz (VAns)\cite{VAns}, both specifically designed to search for the optimal circuit structure.
In addition, we include two further methods in our analysis: Random Adapt-VQE (RA-VQE), an adaptative algorithm which employs a random process to build the circuit structure, and the traditional Quantum Approximate Optimization Algorithm (QAOA), a fixed-structure algorithm.
By analyzing the performance of these methods, we can determine whether an adaptative algorithm can improve the performance of a fixed-structure algorithm and if an adaptative method that utilizes specific ideas for building the circuit structure performs better than an adaptative algorithm that relies on a random approach.

\subsection{Evolutionary Variational Quantum Eigensolver}

Evolutionary Variational Quantum Eigensolver (EVQE) is a genetic algorithm designed to build the quantum circuit for finding the ground state of a Hamiltonian operator. The genomes in the population correspond to different quantum circuits, and their genes correspond to blocks of gates.

EVQE involves several iterations for building the circuits in the population.
At each iteration, the algorithm applies the operations of insertion and removal of gates with fixed probabilities in order to vary the circuit structures while maintaining limited depth.
Differently from other quantum genetic algorithms, EVQE builds the circuits utilizing parametric gates, and optimizes their parameters at each iteration. However, to reduce the computational cost, only the parameters of the last added gene are optimized.
At the end of each iteration, the best circuits are selected and used for building the following generation.
The selection of the best parent genomes is guided by a loss function, defined as follows:
\begin{equation}
\label{loss_EVQE}
\langle \psi|H|\psi \rangle + a \, g + b \, c
\end{equation}
where $\psi$ is the final state of the circuit, $H$ is the Hamiltonian of the problem, $g$ is the total number of genes, $c$ is the number of cnot gates, and $a$, $b$ are two weight coefficients.
This loss is used both to select the best parent genomes at each iteration and to choose the best circuit at the end of the process.
By minimizing the value of the loss function, EVQE is able to identify a circuit with a low expectation and a limited number of gates and cnot gates.

EVQE is different from traditional genetic algorithms because it does not employ crossover (\idest exchange of genes between genomes).
This choice is made because, due to entanglement, merging the circuits of two parent genomes does not necessarily produce an offspring similar to either parent. Therefore, it would be necessary to optimize the parameters of the entire circuit, instead of just the last added gene.

Finally, to preserve population diversification, EVQE employs a speciation strategy, grouping similar individuals into the same species. During the selection of parent genomes, individuals belonging to species with fewer elements are chosen with higher probability, in order to maintain a good representation of all the species.

\subsection{Variable Ansatz}
Variable Ansatz (VAns)\cite{VAns} is an Adaptative VQA, which is chosen because it employs heuristics to guide the selection and placement of gates, as well as simplification strategies to obtain shallower circuits.

VAns starts from an initial layer (``Separable Ansatz'' (SA) or ``Hardware Efficient Ansatz'' (HEA), see Appendix) and performs a number of iterations in which the circuit is modified with the goal of minimizing a loss function. At each iteration, the algorithm inserts blocks of gates compiling to the identity (\idest for which there exists a combination of parameters such that these gates act like the identity gate). The number of blocks to insert is randomly chosen at each iteration according to an exponential distribution. Each block of gates is then applied to randomly selected qubits, with higher probabilities for the qubits with fewer gates.
After inserting gates, the algorithm applies simplification rules that fall into two categories: algebraic and cost-related. Algebraic rules aim to substitute gates with fewer gates that have the same function (\eg simplification of consecutive cnot gates applied to the same qubits or sum of angles in consecutive rotations around the same axis), while cost-related rules remove gates whose contribution to reducing the loss function is negligible.
The algorithm subsequently employs a classical optimizer to adjust the gate parameters (rotation angles), with the objective of minimizing a cost function.
At the end of each iteration, the modifications to the circuit are accepted only if the difference between the new value for the loss and the minimum loss ever found is lower than a threshold.
The stopping criterion is given by a fixed number of iterations of the algorithm.
Further details about VAns can be found in Appendix.

\subsection{Random Adapt-VQE}

In order to provide a comparison between the adaptative algorithms described, which use heuristics to build the circuit structure, with an adaptative algorithm that constructs the circuit through a random process, we introduce a new baseline which we call Random Adapt-VQE (RA-VQE).
We consider two possible choices for the initial layer of the quantum circuit: the same ``Separable Ansatz'' (SA) and ``Hardware Efficient Ansatz'' (HEA), used by VAns (see Appendix).
Then, at each iteration we randomly select a new gate from a pool of rotation and cnot gates, and apply it to randomly chosen qubits. 
After adding a new gate, we optimize all the parameters of the circuit using a classical optimizer.
The algorithm terminates when a certain number of iterations for the search of the circuit structure have been performed.

\subsection{Quantum Approximate Optimization Algorithm}

Finally, in order to provide a comparison with a method that relies on a fixed-structure circuit, we also conduct experiments using the widely adopted Quantum Approximate Optimization Algorithm (QAOA)\cite{QAOA}, a traditional VQA particularly suited to solve combinatorial optimization problems, in particular QUBO problems.
We consider the standard version of QAOA, which includes an initial layer of Hadamard gates, a mixer built with rotation gates around the x-axis, and initial parameters randomly chosen from a uniform distribution.

\section{Experimental Protocol} \label{exp_prot}

In this section we describe our experimental protocol, focusing on the choices we make in order to perform the experiments consistently and compare the different algorithms in terms of quality of the solution found, number of gates and computational time.

\subsection{Problem Instances}

We evaluate the algorithms on three instances: MaxCut, Minimum Vertex Cover, and Number Partitioning, adopting the QUBO formulations described in\cite{glover}. For building the MaxCut instances, we consider both Erdős-Rényi random graphs with an edge probability of 0.7, and, subsequently, a star-topology graph. For Minimum Vertex Cover, we employ other Erdős-Rényi random graphs with an edge probability of 0.7, using a penalty coefficient of 8 in the QUBO formulation described in \cite{glover} (choosing the best value for the penalty is outside the scope of our work). This penalty is applied whenever both endpoints of an edge are excluded from the candidate cover.
Lastly, for Number Partitioning, we generate instances by randomly selecting integer numbers from a list that ranges from 1 to 20.

In each problem, the goal is to find the binary vector that minimizes the QUBO cost function.
In order to find the solution of the QUBO problem, we follow the conventional approach of converting the QUBO formulation into an Ising problem\cite{ising}.
Consequently, the objective becomes to determine the ground state of a Hamiltonian operator, which can be accomplished by utilizing a quantum circuit and minimizing the expectation of the Hamiltonian on its final state.

For each type of problem, we consider instances of different size, with N=4, 8, 12, and 15 variables. For each type of problem and size, we generate ten different QUBO instances, run the algorithms and report the averaged results.

\subsection{Circuit Structure and Parameter Optimization}

The algorithms we consider require two stages. The first stage is the identification of a new candidate circuit structure, which is done according to the specific algorithm criteria (\idest random gate insertion, simplification of redundant gates, selection of best circuits). Once a new candidate structure has been identified, its parameters (\idest angles for the rotation gates) are optimized using a classical optimizer which aims to minimize the circuit expectation. We find the best gate parameters using the SciPy \textit{COBYLA}\cite{cobyla}, a gradient-free and noise-resilient optimizer, largely adopted to tune parameters in variational quantum algorithms\cite{cobyla_performance1, cobyla_performance2}.
Once the optimal parameters for the given structure have been obtained, the structure is evaluated with the specific algorithm loss.
The process is then repeated by generating a new candidate structure, until a stopping criterion is reached.
The choice of the loss function depends on the algorithm being used. For VAns, Random Adapt-VQE, and QAOA, the loss is given by the expectation on the final state, while, for EVQE, it consists in the expectation with added penalties defined in (\ref{loss_EVQE}).
The circuit resulting in the lowest loss function encountered during the optimization process is chosen as the solution for our problem.

\subsection{Hyperparameter Tuning}

\begin{table}[t]
    \caption{Hyperparameters and allowed range for each algorithm\label{hyperp_range}}

    \resizebox{\linewidth}{!}{
    \begin{tabular}{|m{1.8cm} m{1.6cm} | m{2.0cm} m{1.7cm}|}
    \toprule
        EVQE & Range & VAns & Range \\ 
        \midrule
        population size  & \{5, \dots, 20\} & initial layer & \{SA, HEA\}\\ 
        dist threshold  & \{1, \dots, 10\} & scale  & [0, 1.5]\\ 
        prob insertion  & [0, 1]  & temperature  & [1, 20]\\ 
        prob removal  & [0, 1] & accept wall  & [30, 70]\\ 
        a  & [0, 0.5]  & accept perc & [0, 1]\\ 
        b  & [0, 0.5]  & min randomness  & [30, 50]\\
        ~ & ~  & max randomness  & [50, 70] \\  
        ~ & ~  & decrease to  & \{1, \dots, 10\}\\ 
        ~ & ~  & factor accept perc  & [0.8, 0.99]\\         
    \midrule
    \end{tabular}
    }
    
    \resizebox{\linewidth}{!}{
    \begin{tabular}{|p{1.8cm} p{1.6cm} | p{2.0cm} p{1.7cm}|}
        RA-VQE & Range & QAOA & Range \\
        \midrule
        initial layer & \{SA, HEA\} & p & \{1, \dots, 10\} \\ 
    \bottomrule
    \end{tabular}
    }
\end{table}

Each algorithm has a set of hyperparameters that affect the structure search (\eg circuit depth, penalty coefficients, probabilities of adding or removing gates) and need to be optimized to ensure that all methods are evaluated in a fair and consistent way.
For this purpose, we adopt a Bayesian optimization approach\cite{bayesian_opt_tut, bayesian_statistics}, which is widely used to tune hyperparameters in Machine Learning\cite{DBLP:journals/tois/DacremaBCJ21}.
Specifically, for each problem size, we generate a MaxCut instance derived from an Erdős-Rényi random graph with an edge probability of 0.7. Then, for each algorithm and number of variables, we test 50 hyperparameter configurations and choose the one that yields the lowest value for the same loss optimized by the specific algorithm considered.
Since some algorithms are computationally expensive, especially when run on large instances, we set a time limit of 7 days for testing the 50 configurations. If the time limit is exceeded, we select the configuration that leads to the lowest loss among those evaluated.

The complete list of hyperparameters for each algorithm and their ranges of allowed values can be found in Table~\ref{hyperp_range}.
We determine each range by performing some initial trials.

EVQE has many hyperparameters to tune, including the \texttt{population size}, the weights \texttt{a} and \texttt{b} for the penalties in its loss function (\ref{loss_EVQE}), and the probabilities of applying the operations of gate addition (\texttt{prob insertion}) and gate removal (\texttt{prob removal}) at each iteration. The last hyperparameter to tune is the distance threshold \texttt{dist threshold} used to determine if two genomes belong to the same species.

VAns has the highest number of hyperparameters.
The first choice is the \texttt{initial layer}, followed by \texttt{scale}, which regulates the exponential distribution used to determine the number of gates to insert at each iteration and \texttt{temperature}, which defines the probabilities of choosing the qubits where to apply the new gates (a smaller temperature implies that the probability of getting a qubit with fewer gates is higher).
Then, the hyperparameter \texttt{accept wall} determines whether or not to accept the elimination of gates during the cost-related simplification step. \texttt{Accept wall} is varied during the iterations, based on other hyperparameters, referred to as (\texttt{min randomness}, \texttt{max randomness}, and \texttt{decrease to}). 
Finally, the threshold \texttt{accept perc} is used to decide whether to accept a new solution, depending on the difference between its loss and the lowest loss found so far. This threshold is reduced at each iteration using a factor referred to as \texttt{factor accept perc}, which makes the algorithm more selective in accepting worse solutions over time.

For RA-VQE, the only hyperparameter to tune is the \texttt{initial layer}.
QAOA also has only one hyperparameter, which is the depth of the circuit \texttt{p}. However, tuning this hyperparameter is crucial because it affects the expressibility of the circuit, the search space for the optimizer, the computational cost, and the resilience to noise when using real hardware.

\subsection{Computational Budget}

In order to ensure consistency among the different methods, given a problem instance, each algorithm has a maximum budget of $10^4$ evaluations of the circuit expectation: if the maximum budget is reached, the algorithm is stopped.
Then, given a new candidate structure, its parameter optimization has a budget of 50 iterations.
Since QAOA has a fixed-structure circuit, its entire evaluation budget will be used for the parameter tuning phase. Therefore, for QAOA the classical parameter optimizer will have a budget of $10^4$ iterations instead of 50.
We conduct our circuit simulations utilizing the IBM's Qiskit statevector simulator.

\section{Results}

In this section we illustrate the results of our study.
Table~\ref{tab_res_maxcut} shows the results for the execution of all the algorithms on the MaxCut instances (generated both from Erdős-Rényi and star-topology graphs) and Table~\ref{tab_res_vertex_number} shows the results for Vertex Cover and Number Partitioning. 
In particular, we highlight some insights concerning approximation ratio and expectation, number of gates and cnot gates and computational time.

\begin{table*}[t]
    \caption{Results for the MaxCut instances. Approx. Ratio, Expectation, and Time are reported with their respective standard deviations.\label{tab_res_maxcut}}
    \footnotesize
    \resizebox{\linewidth}{!}{
    \begin{tabular}{|cl|ccccc|ccccc|}
    \toprule
            \multirow{3}{*}{N} & 
            \multirow{3}{*}{Algorithm} & 
            \multicolumn{5}{c|}{MaxCut on Erdős-Rényi Graphs}&
            \multicolumn{5}{c|}{MaxCut on Star Graphs}\\

            &
            &
            \begin{tabular}{@{}c@{}}Approx. \\ Ratio\end{tabular} &
            Expectation &
            Gates & 
            Cnot &
            Time [s] &            
            \begin{tabular}{@{}c@{}}Approx. \\ Ratio\end{tabular} &
            Expectation &
            Gates & 
            Cnot &
            Time [s] 
            \\ 
        \midrule
        \multirow{4}{*}{4}  
            & EVQE   & 1.00 ± 0.00 & -3.00 ± 0.94 & 9 & 1 & \phantom{00}231 ± 18\phantom{000} 
                        & 1.00 ± 0.00 & -3.00 ± 0.00 & 10 & 1 & \phantom{00}229 ± 19\phantom{000} \\ 
            & VAns      & 1.00 ± 0.00 & -3.00 ± 0.94 & 2 & 0 & \phantom{00}236 ± 134\phantom{00} 
                        & 1.00 ± 0.00 & -3.00 ± 0.00 & 2 & 0 & \phantom{00}133 ± 30\phantom{000} \\
            & RA-VQE    & 1.00 ± 0.00 & -3.00 ± 0.94 & 6 & 0 & \phantom{00}120 ± 2\phantom{0000}
                        & 1.00 ± 0.00 & -3.00 ± 0.00 & 6 & 1 & \phantom{00}120 ± 2\phantom{0000} \\
            & QAOA      & 0.96 ± 0.09 & -2.85 ± 0.88 & 19 & 7 & \phantom{0000}$\sim 0$\phantom{0000}
                        & 1.00 ± 0.00 & -3.00 ± 0.00 & 17 & 6 & \phantom{0000}$\sim 0$\phantom{0000} \\ 
        \midrule
        \multirow{4}{*}{8} 
            & EVQE   & 0.99 ± 0.01 & -13.98 ± 0.97 & 37 & 5 & \phantom{0}1174 ± 77\phantom{000} 
                        & 1.00 ± 0.01 & -6.97 ± 0.01 & 50 & 7 & \phantom{00}328 ± 20\phantom{000} \\ 
            & VAns      & 0.99 ± 0.04 & -13.90 ± 1.20 & 4 & 0 & \phantom{00}112 ± 14\phantom{000}
                        & 1.00 ± 0.00 & -7.00 ± 0.00 & 5 & 0 & \phantom{00}122 ± 16\phantom{000} \\ 
            & RA-VQE    & 1.00 ± 0.00 & -14.09 ± 0.98 & 16 & 3 & \phantom{00}156 ± 2\phantom{0000}
                        & 1.00 ± 0.00 & -7.00 ± 0.00 & 14 & 2 & \phantom{00}154 ± 4\phantom{0000} \\ 
            & QAOA      & 0.98 ± 0.03 & -13.75 ± 1.01 & 146 & 81 & \phantom{000}35 ± 9\phantom{0000} 
                        & 1.00 ± 0.00 & -7.00 ± 0.00 & 66 & 28 & \phantom{0000}7 ± 1\phantom{0000} \\ 
        \midrule
        \multirow{4}{*}{12} 
            & EVQE   & 0.96 ± 0.01 & -29.83 ± 1.76 & 20 & 1 & \phantom{0}1280 ± 109\phantom{00} 
                            & 0.95 ± 0.02 & -10.47 ± 0.18 & 20 & 1 & \phantom{0}1182 ± 91\phantom{000}\\ 
            & VAns      & 0.95 ± 0.04 & -29.30 ± 2.21 & 5 & 0 & \phantom{00}143 ± 12\phantom{000}
                        & 0.98 ± 0.04 & -10.80 ± 0.42 & 5 & 0 & \phantom{00}210 ± 141\phantom{00} \\ 
            & RA-VQE    & 0.99 ± 0.01 & -30.57 ± 1.55 & 27 & 5 & \phantom{00}731 ± 21\phantom{000} 
                        & 1.00 ± 0.01 & -10.95 ± 0.05 & 20 & 2 & \phantom{00}723 ± 13\phantom{000}\\  
            & QAOA      & 0.98 ± 0.02 & -30.44 ± 1.64 & 471 & 282 & \phantom{0}1020 ± 213\phantom{00}
                        & 1.00 ± 0.00 & -11.00 ± 0.00 & 147 & 66 & \phantom{00}121 ± 9\phantom{0000} \\ 

        \midrule
        \multirow{4}{*}{15}  
            & EVQE   & 0.96 ± 0.02 & -44.13 ± 2.29 & 61 & 7 & \phantom{}14242 ± 1887\phantom{0}
                        & 0.96 ± 0.02 & -13.38 ± 0.34 & 84 & 13 & \phantom{}13296 ± 1500\phantom{0} \\ 
            & VAns      & 0.96 ± 0.03 & -44.60 ± 2.63 & 7 & 0 & \phantom{00}607 ± 90\phantom{000}
                        & 0.99 ± 0.03 & -13.80 ± 0.42 & 7 & 0 & \phantom{0}1553 ± 1360\phantom{0} \\
            & RA-VQE    & 0.98 ± 0.02 & -45.48 ± 1.97 & 29 & 4 & \phantom{}24144 ± 2463\phantom{0}
                        & 0.99 ± 0.01 & -13.88 ± 0.06 & 22 & 2 & \phantom{}24052 ± 1733\phantom{0} \\
            & QAOA      & 0.97 ± 0.02 & -44.60 ± 2.14 & 466 & 281 & \phantom{}24862 ± 3355\phantom{0}
                        & 1.00 ± 0.00 & -14.00 ± 0.00 & 129 & 56 & \phantom{0}3679 ± 471\phantom{00} \\ 
        \bottomrule
    \end{tabular}
    }
\end{table*}

\begin{table*}[t]
    \caption{Results for the Vertex Cover and the Number Partitioning instances. Approx. Ratio, Expectation, and Time are reported with their respective standard deviations. \label{tab_res_vertex_number}}
    \footnotesize
    \resizebox{\linewidth}{!}{
    \begin{tabular}{|cl|ccccc|ccccc|}
    \toprule
            \multirow{3}{*}{N} & 
            \multirow{3}{*}{Algorithm} & 
            \multicolumn{5}{c|}{Vertex Cover on Erdős-Rényi Graphs} &
            \multicolumn{5}{c|}{Number Partitioning}\\
            &
            &
            \begin{tabular}{@{}c@{}}Approx. \\ Ratio\end{tabular} &
            Expectation &
            Gates & 
            Cnot &
            Time [s] &            
            \begin{tabular}{@{}c@{}}Approx. \\ Ratio\end{tabular} &
            Expectation &
            Gates & 
            Cnot &
            Time [s] 
            \\ 
        \midrule

        \multirow{4}{*}{4}  
            & EVQE   & 1.00 ± 0.00 & -25.48 ± 9.08 & 12 & 1 & \phantom{00}214 ± 12\phantom{000}
                        & 1.00 ± 0.00 & -464.09 ± 166.16 & 12 & 1 & \phantom{00}175 ± 12\phantom{000} \\ 
            & VAns      & 1.00 ± 0.01 & -25.40 ± 8.95 & 2 & 0 & \phantom{00}128 ± 22\phantom{000}
                        & 0.99 ± 0.01 & -461.80 ± 167.07 & 2 & 0 & \phantom{00}153 ± 74\phantom{000} \\
            & RA-VQE    & 1.00 ± 0.00 & -25.50 ± 9.09 & 9 & 2 & \phantom{00}121 ± 4\phantom{0000}
                        & 1.00 ± 0.00 & -464.37 ± 166.41 & 10 & 2 & \phantom{00}123 ± 3\phantom{0000} \\ 
            & QAOA      & 0.99 ± 0.01 & -25.20 ± 8.78 & 22 & 7 & \phantom{0000}4 ± 1\phantom{0000}
                        & 1.00 ± 0.00 & -463.64 ± 165.47 & 26 & 12 & \phantom{0000}9 ± 3\phantom{0000} \\

        \midrule
        \multirow{4}{*}{8} 
            & EVQE   & 1.00 ± 0.00 & -153.23 ± 19.08 & 52 & 7 & \phantom{00}318 ± 22\phantom{000}
                        & 1.00 ± 0.00 & -1792.06 ± 810.76 & 38 & 6 & \phantom{00}335 ± 33\phantom{000} \\ 
            & VAns      & 0.99 ± 0.02 & -152.00 ± 17.42 & 6 & 1 & \phantom{00}107 ± 11\phantom{000}
                        & 0.99 ± 0.01 & -1784.50 ± 801.51 & 3 & 1 & \phantom{00}100 ± 9\phantom{0000} \\ 
            & RA-VQE    & 1.00 ± 0.00 & -153.48 ± 19.15 & 14 & 2 & \phantom{00}156 ± 4\phantom{0000}
                        & 1.00 ± 0.00 & -1795.56 ± 811.39 & 13 & 1 & \phantom{00}160 ± 4\phantom{0000} \\  
            & QAOA      & 0.99 ± 0.00 & -152.79 ± 19.17 & 159 & 80 & \phantom{00}242 ± 98\phantom{000}
                        & 1.00 ± 0.00 & -1795.37 ± 810.96 & 192 & 112 & \phantom{00}695 ± 111\phantom{00}\\  
        \midrule
        \multirow{4}{*}{12} 
            & EVQE   & 0.99 ± 0.00 & -356.76 ± 21.04 & 25 & 2 & \phantom{0}1140 ± 67\phantom{000}
                            & 0.99 ± 0.00 & -3751.91 ± 1139.08 & 28 & 3 & \phantom{0}1180 ± 164\phantom{00}\\ 
            & VAns      & 0.97 ± 0.03 & -348.68 ± 21.45 & 10 & 0 & \phantom{00}148 ± 11\phantom{000}
                        & 0.99 ± 0.01 & -3739.40 ± 1139.47 & 4 & 1 & \phantom{00}123 ± 14\phantom{000} \\ 
            & RA-VQE    & 1.00 ± 0.00 & -358.59 ± 20.64 & 21 & 3 & \phantom{00}722 ± 13\phantom{000}
                        & 1.00 ± 0.00 & -3768.12 ± 1141.37 & 20 & 2 & \phantom{00}839 ± 124\phantom{00} \\  
            & QAOA      & 1.00 ± 0.00 & -357.70 ± 20.40 & 498 & 276 & \phantom{0}3825 ± 114\phantom{00}
                        & 1.00 ± 0.00 & -3766.05 ± 1141.98 & 642 & 396 & \phantom{0}4403 ± 167\phantom{00} \\

        \midrule
        \multirow{4}{*}{15}  
            & EVQE   & 0.99 ± 0.00 & -558.47 ± 42.41 & 37 & 4 & \phantom{}14964 ± 3493\phantom{0}
                        & 0.99 ± 0.00 & -5445.82 ± 1150.11 & 35 & 3 & \phantom{}24656 ± 845\phantom{00} \\ 
            & VAns      & 0.99 ± 0.00 & -559.04 ± 42.28 & 14 & 0 & \phantom{0}2287 ± 239\phantom{00}
                        & 1.00 ± 0.00 & -5475.76 ± 1140.00 & 6 & 0 & \phantom{00}813 ± 315\phantom{00} \\
            & RA-VQE    & 1.00 ± 0.00 & -560.98 ± 42.40 & 22 & 2 & \phantom{}20377 ± 4122\phantom{0}
                        & 1.00 ± 0.00 & -5491.80 ± 1158.48 & 26 & 4 & \phantom{}18698 ± 5008\phantom{0} \\
            & QAOA      & 1.00 ± 0.00 & -559.95 ± 42.07 & 505 & 287 & \phantom{}27876 ± 1745\phantom{0}
                        & 1.00 ± 0.00 & -5484.86 ± 1159.61 & 675 & 420 & \phantom{}26206 ± 15692\phantom{} \\ 
        \bottomrule
    \end{tabular}
    }
\end{table*}

\subsection{Approximation Ratio and Expectation}
Our primary focus is the quality of the solution, which we measure using the approximation ratio. This metric is defined as the ratio between the final expectation and the exact cost of the solution.
An approximation ratio close to 1 suggests that the final expectation is close to the cost of the ground state, which in turn indicates a high probability of finding the exact solution when measuring the final state of the circuit. Thus, a method can be considered effective if it leads to an approximation ratio close to 1.

In Table~\ref{tab_res_maxcut} and \ref{tab_res_vertex_number}, we can see that all the algorithms achieve highly accurate solutions with approximation ratios close to 1.

It is worth noting that, while the approximation ratio and the expectation are closely related, we include both in our table because their standard deviations provide different types of information. The standard deviation of the approximation ratio indicates how much the algorithm performance varies across different instances, while the variance of the expectation captures also how the costs of the problem solutions differ from each other.
We observe that the variance of the expectation, when expressed as a percentage, is highest for Number Partitioning. However, the low variance of the approximation ratio for all algorithms suggests that they perform consistently well even when considering different instances.

Moreover, additional experiments (not included here for brevity) confirm that QAOA can achieve good results only when the parameter optimizer is allowed to perform a sufficiently large number of iterations. For instance, limiting the parameter optimizer to 50 iterations results in significantly lower approximation ratios, averaging at 0.96 for N=4, 0.81 for N=8, 0.78 for N=12, and 0.79 for N=15 across the different instances.

\subsection{Number of Gates and Cnot}

Upon the examination of Table~\ref{tab_res_maxcut} and \ref{tab_res_vertex_number}, it is clear that, despite having similar approximation ratios, there is a significant difference between Adaptative VQAs and QAOA in terms of the number of gates and cnot gates in the best circuit found.
The adaptative algorithms require a low number of gates, in particular cnot gates, while QAOA requires a larger number of both.
This represents a significant drawback for QAOA, as it leads to longer computational times for circuit execution and reduces its effectiveness in noisy scenarios.
Moreover, note that in QAOA the number of gates crucially depends on the problem being solved. For example, the star-topology graph has far fewer edges than the Erdős-Rényi random graph, which results in fewer rotation and cnot gates.

Then, it is interesting to note that the random approach discovers a circuit with a relatively low number of gates, even though also deeper circuits are explored (the deepest circuit explored for each instance has around 200 gates). This finding suggests that, even in noise-free scenarios, shorter circuits lead to better results, possibly because the parameter optimizer performs better with fewer angles to tune.

Furthermore, VAns builds circuits with the lowest number of gates and cnot gates, probably because it is the only algorithm that applies simplifications to remove gates, suggesting that the other methods could benefit from a similar technique.
In particular, it is possible to note that, for the instances with N=15, the optimal circuit found by VAns has zero cnot gates.
Overall, VAns exhibits a good balance between solution quality and circuit length, making it a promising approach.

Finally, the algorithms show a significant variation also in circuit depths (not reported due to space limitations). QAOA requires deep circuits, with an average depth often exceeding 100 for instances of size 12 and 15. In contrast, VAns excels in producing shallow circuits, almost always with a depth of 1. EVQE and RA-VQE lie in between, with intermediate circuit depths ranging from 3 to 14.

\subsection{Computational Times}

The time column in Table~\ref{tab_res_maxcut} and \ref{tab_res_vertex_number} indicates the number of seconds for the circuit optimization required by each algorithm and provides valuable information for understanding the runtimes of the different methods.

It is worth noting that VAns almost always achieves the lowest computational time for the instances of size 12 and 15, probably because it tends to create shorter circuits which require less time to execute. Additionally, VAns has the highest variance in computational time, possibly due to the numerous stochastic components that can affect the algorithm runtime.

Moreover, the Adaptative VQAs exhibit similar execution times for different problems (if the number of variables is kept constant). On the other hand, QAOA execution time strongly depends on the specific problem being solved. For example, solving the MaxCut problem on an Erdős-Rényi graph takes at least five times longer than solving MaxCut on a star-topology graph. This discrepancy is likely due to the number of gates in the circuit, which for QAOA is directly tied to the problem being addressed, as observed in the previous subsection.

\subsection{Hyperparameters}

The optimal values obtained for the hyperparameters of each algorithm are reported in Table~\ref{hyperparams_values}.

For EVQE, we can observe that populations with a larger number of individuals are preferred, and that the best coefficients for the penalties in the loss function are almost always 0.
Regarding QAOA, we find that shallower circuits are preferable, even in our noise-free scenario. This is likely because larger values of \texttt{p} lead to circuits with more gates, and therefore a larger number of parameters to tune. As a result, considering deeper circuits may lead to worse solutions because the parameter optimizer struggles to find the global optimum when too many parameters are involved.

Moreover, we remark that selecting appropriate ranges for certain hyperparameters is essential for obtaining a good performance of the algorithm.
For example, \texttt{acceptance wall} in the VAns method must be chosen carefully, otherwise it could cause the removal of all the gates during the cost-related simplification step, leading to an empty circuit and rendering every previous iteration of the VAns algorithm useless.
Similarly, the penalty weights in the loss function of EVQE must be selected with care. If they are set too high, the algorithm may focus only on minimizing the number of gates and ignore the primary objective of optimizing the expectation.

Finally, while experimenting with different hyperparameter configurations, we observed a significant variation in the algorithm loss depending on the chosen configuration. For example, when applied to a MaxCut instance with N=15 variables, VAns achieved a loss value of -49 with the best hyperparameter configuration, while the worst configuration resulted in a loss of -37.

\begin{table}[t]
    \caption{Best hyperparameters found with the Bayesian optimizer \label{hyperparams_values}}
    \resizebox{\linewidth}{!}{
    \begin{tabular}{|l|lllll|}
    \toprule
        Algorithm & Hyperparameter & N=4 & N=8 & N=12 & N=15\\ 
        \midrule
        \multirow{6}{*}{EVQE}
        & population size & 20 & 19 & 20 & 14 \\ 
        & dist threshold & 1 & 7 & 4 & 3 \\ 
        & prob insertion & 0.282 & 0.451 & 0.164 & 1.000 \\ 
        & prob removal & 1.000 & 0.000 & 0.287 & 0.178 \\ 
        & a & 0.000 & 0.000 & 0.000 & 0.000 \\ 
        & b & 0.000 & 0.000 & 0.098 & 0.000 \\ 

        \midrule 
        \multirow{10}{*}{VAns}
        & initial layer & SA & SA & SA & SA\\ 
        & scale & 1.137 & 0.459 & 0.489 & 0.000 \\ 
        & temperature & 11.477 & 8.027 & 5.737 & 20.000 \\ 
        & accept wall & 58.522 & 55.661 & 34.411 & 30.000 \\ 
        & accept perc & 0.042 & 0.323 & 0.096 & 0.000 \\ 
        & min randomness & 49.968 & 39.338 & 38.776 & 50.000 \\ 
        & max randomness & 67.402 & 51.267 & 51.222 & 70.000 \\ 
        & decrease to & 7 & 9 & 1 & 10 \\ 
        & factor accept perc & 0.820 & 0.810 & 0.973 & 0.800 \\ 

        \midrule 
        \multirow{1}{*}{RA-VQE} & initial layer & SA & SA & SA & SA \\ 

        \midrule 
        \multirow{1}{*}{QAOA}
        & p & 1 & 2 & 3 & 2 \\ 
        \bottomrule
    \end{tabular}
    }
\end{table}

\section{Conclusions}

Our study benchmarks three different Adaptative Variational Quantum Algorithms (EVQE, VAns, and RA-VQE) and QAOA for solving QUBO problems in noise-free scenarios.
Our results show that the adaptative algorithms do not provide an advantage in terms of approximation ratio, since the value obtained with all methods is always close to 1.
However, we observe significant differences in the number of gates and runtimes. QAOA generates high-quality solutions, but its fixed-structure circuit requires a high number of gates, increasing the computational time and making it more prone to errors in noisy scenarios. On the other hand, Adaptative VQAs, in particular VAns, can find shorter circuits that maintain solution quality, making them promising alternatives to traditional VQAs.
Similarly, from the comparison between EVQE and VAns with RA-VQE, the former do not show an advantage in terms of approximation ratio, but they sometimes allow to obtain a shorter circuit.
Moreover, it is worth noting that the adaptative algorithms offer the significant advantage of producing flexible circuits, which can be tailored to the constraints and requirements of the specific hardware.

Finally, our work suggests potential directions for future research. One possible avenue is the investigation of the performance of adaptative algorithms on larger problem instances and in noisy scenarios, as well as the evaluation of their effectiveness on real quantum hardware.
Furthermore, the techniques adopted by a particular adaptative algorithm, such as the circuit simplification strategies employed by VAns, could potentially be adapted for use in other algorithms.

Overall, our study provides insights into the potential of Adaptative VQAs and emphasizes the importance of considering all relevant dimensions when evaluating these algorithms: circuits leading to solutions with similar approximation ratios can have a significantly different number of gates, which is a crucial dimension to consider when assessing the efficiency and feasibility of quantum algorithms.

\appendix

Here we specify some details and specific choices made for the VAns algorithm\footnote{Some specific choices are not included in the paper, but based on the GitHub code available at \url{https://github.com/matibilkis/qvans} and on a discussion with the authors.}.
Firstly, we consider the ``Separable Ansatz'' (SA) and ``Hardware Efficient Ansatz'' (HEA) in FIG. 2 of the original paper\cite{VAns} as options for the initial layer, and use the blocks of gates depicted in FIG. 3 as options for the gates to insert at each iteration.

The circuit simplification process involves applying algebraic rules (Section IIC, rules 1 to 5) iteratively until no further modifications are possible. Then, we use the commutation rules from FIG. 4 and check if the circuit can be further simplified applying the algebraic rules again. These simplification and commutation rules are applied iteratively until no further simplification is possible.
Next, we perform cost-related simplifications by removing each gate from the circuit, one at a time, and evaluating the expectation without that gate. If this expectation is lower than the expectation obtained considering the full circuit, or higher but within a certain threshold, we consider the modified circuit as acceptable. After testing the removal of all the gates, we choose the circuit leading to the lowest expectation among the acceptable ones. We continue this process of simplification iteratively, until no removal is considered acceptable.
We then repeat the process of applying both algebraic and cost-related simplifications, until no further modification is possible.

Finally, it is worth noting that the probabilities for choosing the number of blocks of gates to insert at each iteration and the qubits where to place them, as well as the formulas for varying the hyperparameters \texttt{temperature}, \texttt{accept wall} and \texttt{accept perc} at the end of each iteration are derived from the GitHub code with the original implementation of VAns.

\FloatBarrier 

\clearpage

\section*{Acknowledgments}
We acknowledge the financial support from ICSC - ``National Research Centre in High Performance Computing, Big Data and Quantum Computing'', funded by European Union – NextGenerationEU.

\bibliographystyle{IEEEtran}
\bibliography{IEEEabrv,ms}

\end{document}